\documentclass[conference]{IEEEtran}
\IEEEoverridecommandlockouts
\usepackage{cite}
\usepackage{amsmath,amssymb,amsfonts}
\usepackage{algorithmic}
\usepackage{graphicx}
\usepackage{textcomp}
\usepackage{xcolor}
\usepackage{tikz}
\usepackage[font={footnotesize}]{subcaption}
\usepackage[font={footnotesize}]{caption}
\usepackage{import}
\usepackage{enumitem}
\usepackage{dirtytalk}

\renewcommand{\theenumii}{\theenumi.\arabic{enumii})}

\usepackage{tabularx,booktabs}
\newcolumntype{C}{>{\centering\arraybackslash}X} 
\usepackage[official]{eurosym}
\usepackage{orcidlink}
\usepackage{gensymb}

\usepackage{graphicx,wrapfig,lipsum}
\def\BibTeX{{\rm B\kern-.05em{\sc i\kern-.025em b}\kern-.08em
    T\kern-.1667em\lower.7ex\hbox{E}\kern-.125emX}}

\makeatletter

\let\old@ps@IEEEtitlepagestyle\ps@IEEEtitlepagestyle
\def\confheader#1{%
    \def\ps@IEEEtitlepagestyle{%
        \old@ps@IEEEtitlepagestyle%
        \def\@oddhead{\strut\hfill#1\hfill\strut}%
        \def\@evenhead{\strut\hfill#1\hfill\strut}%
    }%
    \ps@headings%
}

\makeatother

\confheader{%
    \parbox{20cm}{\textbf{\textit{Accepted to IEEE Belgrade PowerTech, 2023}}}
}

\begin{document}

\title{End-to-End Learning with Multiple Modalities for System-Optimised Renewables Nowcasting}

\thispagestyle{plain}
\pagestyle{plain}

\author{\IEEEauthorblockN{Rushil~Vohra,
        Ali~Rajaei\orcidlink{0000-0002-1944-546X},~\IEEEmembership{Graduate Student Member,~IEEE,} and~Jochen~L.~Cremer \orcidlink{0000-0001-9284-5083}~\IEEEmembership{Member,~IEEE,}}
        
\IEEEauthorblockA{\emph{Department of Electrical Sustainable Energy} \\
\emph{Delft University of Technology} \\
Delft, The Netherlands
    \\rushilvohra@outlook.com, \{A.Rajaei, J.L.Cremer\}@tudelft.nl}
}

\maketitle
\vspace{-4mm}
\begin{abstract}
With the increasing penetration of renewable power sources such as wind and solar, accurate short-term, nowcasting renewable power prediction is becoming increasingly important. This paper investigates the multi-modal (MM) learning and end-to-end (E2E) learning for nowcasting renewable power as an intermediate to energy management systems. MM combines features from all-sky imagery and meteorological sensor data as two modalities to predict renewable power generation that otherwise could not be combined effectively. The combined, predicted values are then input to a differentiable optimal power flow (OPF) formulation simulating the energy management. For the first time, MM is combined with E2E training of the model that minimises the expected total system cost. The case study tests the proposed methodology on the real sky and meteorological data from the Netherlands. In our study, the proposed MM-E2E model reduced system cost by $30$\% compared to uni-modal baselines.

\end{abstract}

\vspace{-2mm}
\section{Introduction}
\label{sec:intro}

In many regions, sustainable energy systems will rely heavily on photovoltaic (PV) power, wind power, or both. These forms of energy generation are inherently uncontrollable due to their dependence on the local weather. Therefore, predicting energy generation is essential to ensure a balanced electrical system. A shared goal of system operators and microgrid managers is maintaining grid stability cost-effectively, where forecasting power can be considered an intermediate step to managing the energy system.

Conventional numerical weather prediction (NWP) models and statistical methods are for instance autoregressive moving average (ARMA) and autoregressive integrated moving average (ARIMA). However, these NWP and statistical models sometimes result in inaccurate forecasts, especially in the short term. Applying machine learning (ML) methods for these predictions in recent years aimed at lowering these inaccuracies. For example, models based on artificial neural networks (ANNs) improved prediction accuracy using meteorological time series input data based on input features such as humidity and air pressure for forecasting outputs such as solar irradiance and wind speed \cite{REN201582,Ahmed2020}. Authors in \cite{8810672} developed a hybrid PV power short-term prediction method that combines an ARIMA model with an ANN model and considers the different seasons as model inputs. Reference \cite{8810921} proposed an approach that considers meteorological inputs and produces ensemble predictions. Beyond meteorological data, \cite{zhang2018deep} uses another type of ANN, a convolutional neural network (CNN), and a long short and term memory (LSTM) module that takes sky images as input to produce the expected PV power as output. Despite this impressive development improving the accuracy over the years through machine learning, ML-based approaches for PV and wind predictions do not consider heterogeneous input data simultaneously in their predictions. As these data sources, meteorological and sky images, contain different information for predicting PV and wind output, their combination is promising. 

Multi-modal (MM) learning combines time series data with imagery and improves the prediction of severe short-term fluctuations that other ANN approaches may miss \cite{Li2019}. MM learning combines different modalities of heterogenous data streams (such as text and audio) so that features from either modality enhance each other \cite{Baltrusaitis2019}. Combining these modalities means learning a common function for all the modalities simultaneously to produce a single representation vector. This `function' can be an ANN. Three different 'fusion' methods to combine these modalities are early, late, and intermediate fusion\cite{Lahat2015MultimodalDF}. Early fusion combines raw or pre-processed data from different modalities, e.g., by concatenation, before the input to an ML model. Late fusion combines the predictions made from either modality, e.g., using a weighted average. Finally, intermediate fusion implements a hidden layer, called as MM layer, in the ANN. The engineer designs the ANN model architecture and selects this layer's position in the ANN. 
Recently, \cite{Li2019} showed that MM learning successfully forecasted PV power outperforming baselines by a large margin. Beyond power prediction, MM also improved energy load time series forecasting \cite{10.1007/978-981-19-2456-9_115}.

End-to-end (E2E) learning models the task of the user in the training of the ML model \cite{Donti2017}. This learning differs from conventional learning of ML models that uses a prepared (labelled) data set and only assumes a loss on the training data. This assumption has a weakness of not considering the task for what the trained ML model has used afterwards, which is what E2E learning addresses. For example, an ML model trained for predicting PV or wind power in power systems could have the final task of being used as input optimising an energy management schedule of resource allocations \cite{Dar23}. In more detail, from the perspective of a power system operator, the task can be modelled as a mathematical optimisation problem with the objective to minimise the total system operational cost and constraints that ensure the grid's physical security limits. 
Donti et al. \cite{Donti2021} developed an E2E learning approach for convex optimisations such as the optimal power flow (OPF) problem. E2E learning considers then the optimisation, here an OPF,  that can be modelled in an ANN as a differentiable optimisation layer making use of the implicit function theorem \cite{cvxpylayers2019}. The ANN is then trained on the system's true objective, the economic cost \cite{Donti2021}. 

This paper combines MM with E2E for the first time and applies our \textit{MM-E2E} technique to a MM PV and wind forecasting using an ANN that is trained E2E based on the system's cost arising from inaccurate PV and wind forecasts. Our proposed MM ANN combines features with intermediate fusion from imagery and time-series meteorological data to predict PV and wind generation. Respectively, the predicted PV and wind generation are input to a differentiable optimisation layer that determines the schedule to dispatch power generations cost-efficiently and minimises their re-dispatch cost. The two contributions are:
\begin{enumerate}[label=(\roman*)]
    \item \textit{MM-E2E} combines the intermediate fusion of MM and E2E learning of a model for PV and wind generation prediction in energy systems, leveraging their advantages: MM maximises learnable information from heterogeneous data, and E2E learns models for accurate predictions.
    \item E2E learning for an energy system with PV and wind generation. The dispatching and redispatching optimisation problem captures the ultimate system economic cost caused by prediction inaccuracies. The ANN model learns to compensate for the inaccuracies in the most cost-effective way.
\end{enumerate}
The rest of this paper is structured as follows. Section \ref{sec:method} outlines the design and formulation of the proposed \textit{MM-E2E} model. The case study in Section \ref{sec:cases} presents the numerical results of applying the proposed framework on the IEEE $6$-bus system, followed by a discussion. The paper is concluded in Section \ref{sec:conc}.
\vspace{-2mm}
\section{Methodology}
\label{sec:method}
\par This section describes the proposed \textit{MM-E2E} learning approach for solar and wind prediction along the workflow of three steps: pre-processing, MM learning and E2E with optimal power flow as task optimisation. After pre-processing the heterogeneous data sources, i.e., imagery and meteorological sensor data, the proposed approach combines these data features with MM. Subsequently,  E2E learning trains the ANN-based model on the ultimate economic system cost arising from prediction errors. 

\subsection{Data Pre-processing}
To reduce the computational times of image processing, the imagery dataset is pre-processed in two ways. Firstly, the images are converted from colour scale to grayscale, which may also benefit object recognition \cite{Bui2016}. Secondly, the resolution of the imagery is reduced from $1536 \times 1536$ to $64 \times 64$. 
Third, as ANNs training converges faster with input feature values between $0$ and $1$, the image pixel values are min-max normalised  \cite{robotics}. Hence the resulting images $i \in \Omega^N$ have the pixel features $x_i^{Image} \in (0,1)^{64 \times 64}$. The tensor $X$ combines many of such images, e.g., $X^{Image} \in (0,1)^{64 \times 64 \times N}$ where $N$ is the cardinality $|\Omega^{N}| = N$ of all (training) images $\Omega^{N}$. Also, min-max normalises the meteorological data resulting in $x_i^{Meteo} \in (0,1)^{F \times N}$ with the number of meteorological features $F$. Finally, imagery and meteorological data samples were measured every $10$ minute to nowcast renewable power from $10$-minutes-recent data. 

\subsection{Multi-Modal Learning: Intermediate Fusion of Features}
Convolutional neural network (CNN) $f_1$ can predict PV or wind power $\hat{y}_i \in \mathbb{R}_{\geq 0}^{2}$ exclusively from imagery $x_i^{Image}$ and feedforward NN (FNN) $f_2$ 
is suitable for exclusively using numerical (meteorological) data $x_i^{Meteo}$, respectively
\begin{subequations} \label{rawANNs}
\begin{align}
f_1: x_i^{Image} \rightarrow \hat{y}_i \\
f_2 : x_i^{Meteo} \rightarrow \hat{y}_i.
\end{align} 
\end{subequations}

A typical CNN architecture contains stacks of convolutional layers with activation functions, followed by a pooling layer and, optionally, a batch normalisation layer \cite{robotics}. Fully-connected (FC) layers with a specified activation function are stacked 'on top' to make the final predictions $\hat{y}_i$. 

To predict power combined from the two modalities of data, feature information from either one must be considered in conjunction. \textit{MM-E2E} via intermediate fusion conjuncts these pieces of information: features $h_{1,i}$ are extracted from the imagery using a CNN 
\begin{equation}
\tilde{f}_1: x_i^{Image} \rightarrow h_{1,i} \label{eq:CNN-fe} 
\end{equation}
and concatenated with features from the meteorological data
\begin{equation}
h_{mm,i} = {h_{1,i}}^\frown x_i^{Meteo}. \label{eq:concatenation} 
\end{equation}

This intermediate fusion of features was selected as the early fusion of input data directly ${x_i^{Image}}^\frown x_i^{Meteo}$ is not suited. Fusing early imagery and sensor data as modalities without any feature extraction is not well suited as the sensor data cannot be directly concatenated with the pixel values of an image, and different types of ANNs are suited for the different data types. 
Another suggested approach for two modalities is bilinear pooling \cite{morency2017} with the outer product
\begin{equation}
    \label{eq:bipool}
    h_{mm} = \begin{bmatrix}
                h_1 \\
                1
             \end{bmatrix} \otimes
             \begin{bmatrix}
                h_2\\
                1
             \end{bmatrix} = 
             \begin{bmatrix}
                h_1 & h_1 \otimes h_2 \\
                1 & h_2
             \end{bmatrix},
\end{equation}
where the index $i$ was dropped for simplicity reasons for the two feature tensors $h_1$ and $h_2$ that give the MM feature tensor $h_{mm}$, where $h_{1,i}$ is the extracted feature tensor for the first modality, and $h_{2,i}$ is the extracted feature tensor for the second.  Subsequently, another ANN 
\begin{equation}
f_3 : h_{mm,i} \rightarrow \hat{y}_i, \label{eq:finalANN}
\end{equation}
learns predicting the PV and wind power levels $\hat{y}_i = (\hat{P}^{PV}_i, \hat{P}^{W}_i)$ from the concatenated, extracted features $h_{mm}$.

\par A possible disadvantage of intermediate fusion is that the approach may result in a very large and imbalanced feature tensor. However, it has also been argued that these imbalances in the tensor accelerate the convergence of training the ANN. The case study, Section \ref{case7} compares these two approaches.

\subsection{Optimal Power Flow} \label{sec:OPF}
After the ANN $f_3$ predicts renewable power generation $\hat{y}_i$, two varying OPF optimisation problems are solved in a sequence minimising the system cost and ensuring the physical constraints of the grid. The first problem referred to as \textit{OPF-Schedule}, receives the predicted renewable generation $\hat{y}_i$ from the ANN $f_3$ as a parameter and aims to minimise system cost by scheduling the dispatch of generators. However, during the real-time operation, the predicted renewable generation may be different from the true value, which may lead to energy imbalances in the system. Therefore, the second optimisation problem, referred to as \textit{OPF-Redispatch}, is formulated to find the optimal redispatch of the generators that can balance supply and demand with the lowest redispatch cost. 
 
The transmission grid consists of a set of nodes $B$, and a set of lines $L$. Without loss of generality, it is considered that each node $b \in B$ could be connected to a conventional generator, load demand, PV and wind generation. Therefore, ANN $f_3$ outputs the PV and wind power predictions at multiple locations, and we expand $\hat{y}_{i,b} = (\hat{P}^{PV}_{i,b}, \hat{P}^{W}_{i,b})$ by the index $b$ for the bus locations $\forall b \in B$. Then, we drop the index $i$ in the following for simplicity reasons and continue only with $\hat{P}^{PV}_b$ ($\hat{P}^{W}_b$). The \textit{OPF-Schedule} assumes the DC power flow equations with known load demand as follows
\begin{eqnarray} \min &C^{sch} = \sum_{b\in B} c_b^g P_b^g  + \sum_{b\in B} \big( \gamma^{PV}_b\lambda^{PV}_b \nonumber \\  & +\gamma^{W}_b\lambda^{W}_b  + \gamma_b(\lambda^+_b + \lambda^-_b) \big)
 \label{eq:schcost}
\end{eqnarray}
\begin{eqnarray} \textrm{s.t.}& P^g_b-P^d_b+(\hat{P}^{PV}_b-\lambda^{PV}_b)+(\hat{P}^{W}_b-\lambda^{W}_b) \nonumber \\
& \quad = \sum_{l\in L_b} f_l +(\lambda^{+}_{b}-\lambda^{-}_{b})\quad \quad \forall b \in B      \label{eq:nodebalance} 
\end{eqnarray}
\begin{eqnarray} & \underline{P^g_b}  \leq P_b^g \leq \overline{P^g_b}   \quad \quad \forall b \in B \label{eq:genlim}
\end{eqnarray}
\begin{eqnarray} & f_l = \frac{\delta_{i}-\delta_{j}}{x_l} \quad \quad \forall l \in L \label{eq:branchflow}
\end{eqnarray}
\begin{eqnarray} & -\overline{f_l} \leq f_l \leq \overline{f_l}  \quad \quad \forall l \in L 
\label{eq:thermal}
\end{eqnarray}
\begin{eqnarray} & 0 \leq \lambda^{PV}_b \leq \hat{P}^{PV}_b, \quad 0 \leq \lambda^{W}_b \leq \hat{P}^{W}_b \quad \forall b \in B  \label{eq:lamPV}
\end{eqnarray}
where the objective function in \eqref{eq:schcost} minimises the operational dispatching cost of the system denoted by $C^{sch}$, which consists of dispatching cost, renewable curtailment cost, and a penalty term for infeasibility in the node balance. In this regard, the conventional generator output and cost are $P^g_b$ and $c^g_b$. Furthermore, the predicted PV (wind) generation is $\hat{P}^{PV}_b$ ($\hat{P}^{W}_b$), while PV (wind) curtailment and corresponding curtailment cost are $\lambda^{PV}_b$ ($\lambda^{W}_b$), and $\gamma^{PV}_b$ ($\gamma^{W}_b$), respectively. Positive (negative) node injection imbalance and its associated penalty are given by $\lambda^+_b$ ($\lambda^-_b$) and $\gamma_b$. Furthermore, load demand and active line flow are represented by $P^d_b$, and $f_l$, respectively.  Eq. \eqref{eq:nodebalance} models the node balance, while the flow of line $l$ between adjacent nodes with the reactance of $x_l$ is formulated in \eqref{eq:branchflow}. Finally, the limits on generator output, line flow, PV and wind curtailment are presented in \eqref{eq:genlim}, \eqref{eq:thermal}-\eqref{eq:lamPV}.      
While the dispatch of generators $\hat{P}^g_b$ is determined based on the predicted renewable generation $\hat{P}^{PV}_b$ and $\hat{P}^{W}_b$, the true values of renewable generation  $P^{PV}_b$ and $P^{W}_b$, might be different, e.g. 
\begin{subequations}
\begin{align}\label{eq:mismatch}
& \hat{P}^{PV}_b \neq P^{PV}_b && \exists \, b \in B \\
&\hat{P}^{W}_b \neq P^{W}_b &&\exists \, b \in B. 
\end{align}
\end{subequations}
Therefore, redispatching accounts for this mismatch between the predicted and actual renewable generation. In this context, the \textit{OPF-Redispatch} problem is formulated as below:
\begin{eqnarray} \min &C^{rd} = \sum_{b\in B}(c_b^{up}dP_b^{g+} + c_b^{dn}P_b^{g-})  \nonumber \\
&  + \gamma^{PV}_b\lambda^{PV}_b  + \gamma^{W}_b\lambda^{W}_b + \sum_{b\in B} \big( \gamma_b(\lambda^+_b + \lambda^-_b) \big)
 \label{eq:rdcost}
\end{eqnarray}
\begin{eqnarray} \textrm{s.t.}& (\hat{P}^g_b+ dP_b^{g+} - dP_b^{g-}) -P^d_b  \nonumber \\ 
&+(P^{PV}_b-\lambda^{PV}_b) + (P^{W}_b-\lambda^{W}_b) \nonumber \\ 
& =\sum_{l\in L_b} f_l +(\lambda^{+}_{b}-\lambda^{-}_{b}) \quad \quad \forall b \in B \label{eq:rdbalance}
\end{eqnarray}
\begin{eqnarray}
0 \leq dP_b^{g+} \leq \overline{dP_b^{g+}} , \quad
0 \leq dP_b^{g-} \leq \overline{dP_b^{g-}} \quad \quad \forall b \in B \label{eq:rdlimit}
\end{eqnarray}
\begin{eqnarray} & \underline{P^g_b}  \leq \hat{P}_b^g+ dP_b^{g+} - dP_b^{g-}  \leq \overline{P^g_b} \quad \quad \forall b \in B \label{eq:rdgenlimit}
\end{eqnarray}
\begin{eqnarray} & \quad \text{Eqs. } \eqref{eq:branchflow}-\eqref{eq:lamPV}, \nonumber
\end{eqnarray}
where the objective minimises the redispatching cost of the system, shown by $C^{rd}$, which includes upward and downward ramping cost of generators, as well as renewable and penalised nodal imbalances. In this regard, upward (downward) redispatch of generators and its associated cost is represented by $dP_b^{g+}$ ($dP_b^{g-}$), and $c_b^{up}$ ($c_b^{up}$), respectively. The nodal balance constraint is shown in \eqref{eq:rdbalance}, while generator redispatch constraints are presented in \eqref{eq:rdlimit}-\eqref{eq:rdgenlimit}. 
\par After solving OPF-Redispatch, the total system cost $C^{sys}$ can be calculated as below:
\begin{eqnarray} \label{eq:systemcost}
	& C^{sys} = C^{sch} + C^{rd}
\end{eqnarray}

The total system cost is a function of renewable prediction accuracy, where prediction errors lead to higher operational costs $C^{sys}$, where the root-mean-squared error for PV and wind are Eq. \eqref{RMSEPV} and Eq. \eqref{RMSEW} respectively.
\begin{equation}
RMSE^{PV}= \sqrt{ \frac{\sum_{b\in B} (\hat{P}^{PV}_b - {P}^{PV}_b)^2} {|B|}} \label{RMSEPV}
\end{equation}
\begin{equation}
RMSE^{W}= \sqrt{ \frac{\sum_{b\in B} (\hat{P}^{W}_b - {P}^{W}_b)^2}{|B|}}
\label{RMSEW}
\end{equation}

\subsection{Proposed Multi-Modal End-to-End Learning Approach}
The proposed \textit{MM-E2E} learning approach trains the ANNs $\tilde{f_1}$ and $f_3$ to minimise the ultimate system cost $C^{sys}$. Our approach combines the features from the modalities $h_1$ and $x^{Meteo}_i$ in a single ANN $f_3$ to perform renewable power prediction, which is then used as input for OPF optimisation as the ultimate task. The proposed approach is shown in Fig. \ref{fig:mm4} and follows $5$ consecutive steps:

\begin{figure}
    \centering
    \includegraphics[width=0.5\textwidth]{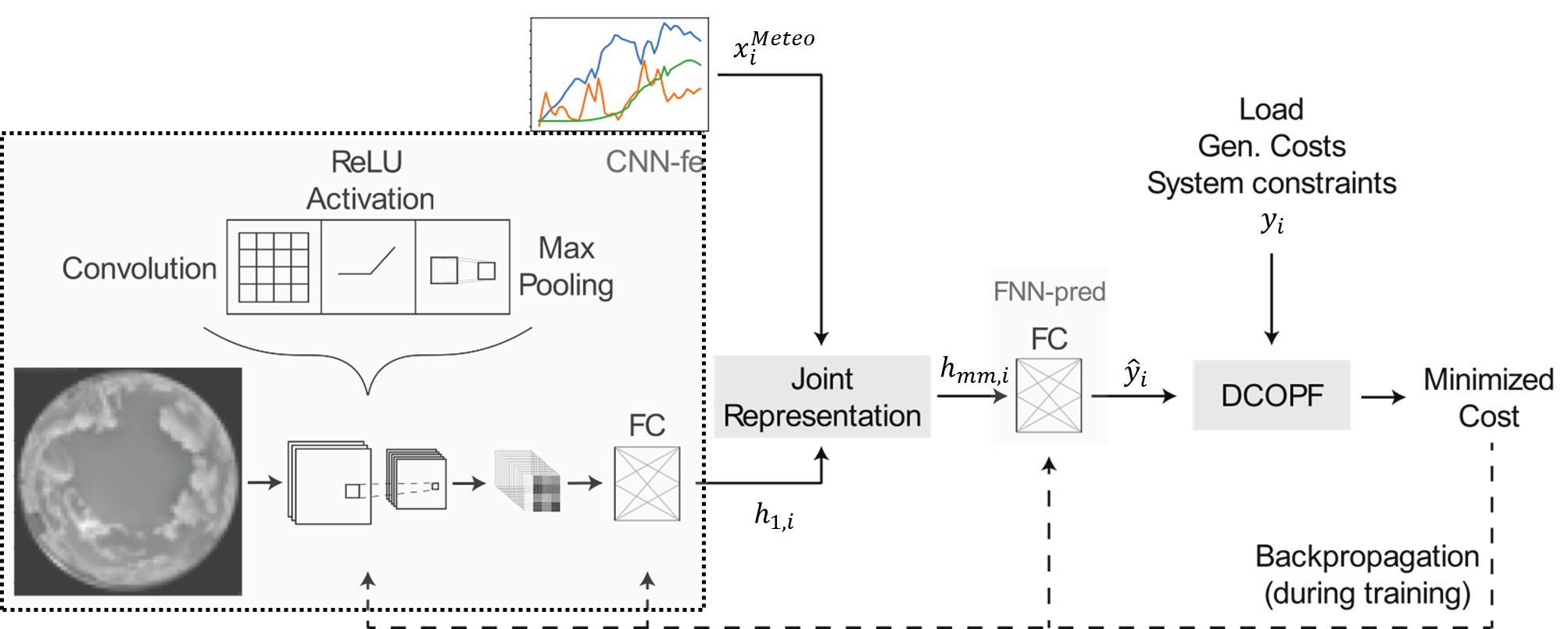}
    \caption{The proposed multi-modal end-to-end architecture (\textit{MM-E2E}). The extracted features from imagery are fused with the meteorological data to predict renewable power ($\hat{y}$) by FC layers. The power prediction is then used to solve the OPF problems, the output of which trains the ANN.}
    \label{fig:mm4}
    \vspace{-5mm}
\end{figure}

\begin{figure}
    \centering
    \includegraphics[width=0.49\textwidth]{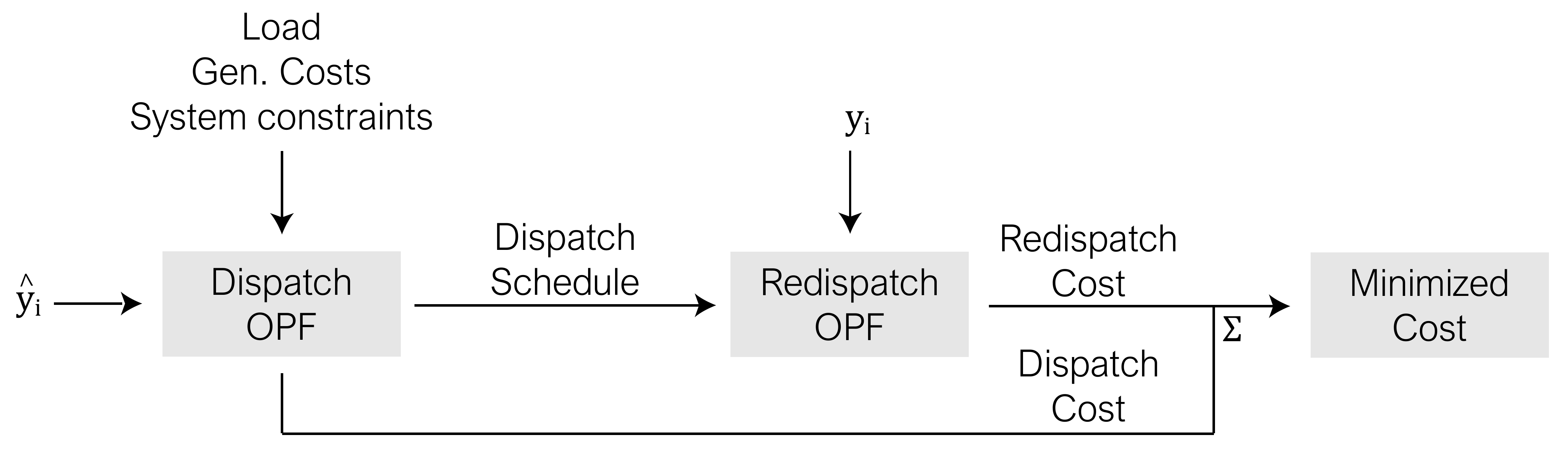}
    \caption{The \textit{OPF-Schedule} and \textit{OPF-Redispatch} minimise system costs.}
    \label{fig:opf}
\vspace{-6mm} 
\end{figure}
\begin{enumerate}
    \item A feature extraction network $\tilde{f_1}$, \textit{CNN-fe} is applied to an image input $x^{Image}_i$ according to Eq. \eqref{eq:CNN-fe}. The \textit{CNN-fe} shown as part of Fig. \ref{fig:mm4} consists of convolutional, ReLU activation, and max pooling layers, followed by an FC layer with ReLU activation and another with Sigmoid activation. $\tilde{f_1}$outputs the feature vector $h_{1,i}$. 
    \item The meteorological features $x^{Meteo}_i$ 
    are fused with $h_{1,i}$ in a joint representation using concatenation as in Eq. \eqref{eq:concatenation}
    \item Renewable power $\hat{y}_i$ is predicted from the fused feature vector, $h_{mm,i}$, using a FNN layer $f_3$ as in Eq. \eqref{eq:finalANN}, referred to as \textit{FNN-pred}. 
    \item The prediction $\hat{y}_i=(\hat{P}^{PV}_i,\hat{P}^{W}_i)$ is one of the inputs to the OPFs as illustrated in Fig. \ref{fig:opf}. These OPFs consist of two successively executed variations of OPFs (\textit{OPF-Schedule} and \textit{OPF-Redispatch}) as in Sec. \ref{sec:OPF}. Both are integrated as differentiable parameterized optimization layers in the neural network, where the solutions are differentiated concerning the parameters $\hat{y}_i$ \cite{cvxpylayers2019}.
    \begin{enumerate}
    \item As shown in Fig. \ref{fig:opf}, the renewable prediction $\hat{y}_i=(\hat{P}^{PV}_i,\hat{P}^{W}_i)$ is the input to the optimal dispatch schedule of the generators with \textit{OPF-Schedule}.
    \item Then, if there exists any power imbalance due to a mismatch of the predicted power ($\hat{P}^{PV}_b,\hat{P}^{W}_b)$ to the true powers $(P^{PV}_b,P^{W}_b)$ according to Eq. \eqref{eq:mismatch}, the \textit{OPF-Redispatch} problem is solved to find the optimal redispatch of the generators. 
    \end{enumerate}
    \item During training, the loss is considered as the total system cost of dispatch and redispatch, given by Eq. \eqref{eq:systemcost}. This loss is differentiated and the gradients backpropagated to update the weights of all parameters defining functions $\tilde{f_1}$ and ${f_3}$ of \textit{CNN-fe} and \textit{FNN-pred}, respectively.
\end{enumerate}
\begin{figure}
    \centering
    \includegraphics[width=0.3\textwidth]{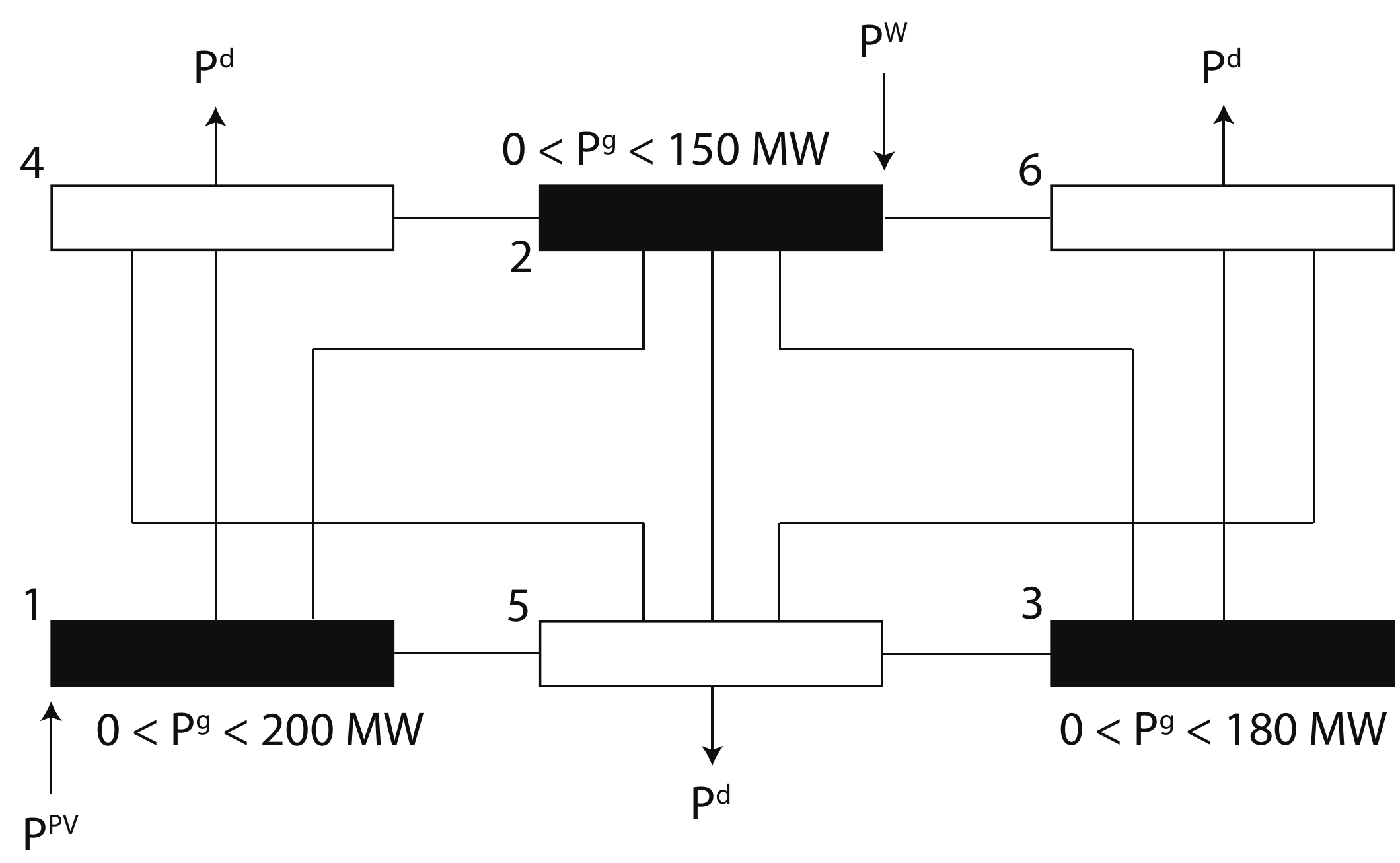}
    \caption{The modified IEEE 6-bus system with PV and wind generation.}
    \label{fig:line2}
    \vspace{-5mm}
\end{figure}

\section{Case Studies}
\label{sec:cases}
\subsection{Settings and Test Network}
\subsubsection{Power System}
Fig. \ref{fig:line2} shows the modified IEEE $6$-bus system with $3$ generator, $3$ load demands, PV generation at bus $1$, and wind generation at bus $2$. Other parameters used in \textit{OPF-Schedule} and \textit{OPF-Redispatch} optimisations are presented in Appendix \ref{appendix}. Below are details regarding load demand, PV generation, and wind generation. 
\begin{itemize}
	\item The load is modelled from the dynamic load profiles of $1.5$ million identical households, normalised to IEEE $6$-bus values and distributed over $3$ buses \cite{bdew}. In this context, the maximum load demand of the considered data is $293$MW. 
	\item PV generation is based on the LG NeON BiFacial 440W panel (with a panel area of $2.2$ m$^2$). Eq. \eqref{eq:pvoutput} approximates the output of a PV farm with $N_{panels}$ panels of area $A_{panel}$, given their rated power $P_{rated}$ and the Global Horizontal Irradiance (GHI).
    \begin{equation} \label{eq:pvoutput}
        P_{PV} = \frac{GHI}{1000 W/m^2}\times P_{Rated}\times N_{panels}
    \end{equation}
    Eq. \eqref{eq:pvoutput} disregards losses in panel efficiency due to temperature. We assume a farm of $250,000$ panels and corresponding rating of $110$MWp. 
	\item Wind generation is based on a farm of 36 Vestas V112-3.45MW turbines (modelled on the existing OWEZ offshore farm), resulting in a rating of 124MWp. The turbine power curve is approximated by a Tanh function, which can be found in Appendix \ref{appendix}.
\end{itemize}
\par
\begin{figure}
    \centering
    \includegraphics[width=0.2\textwidth]{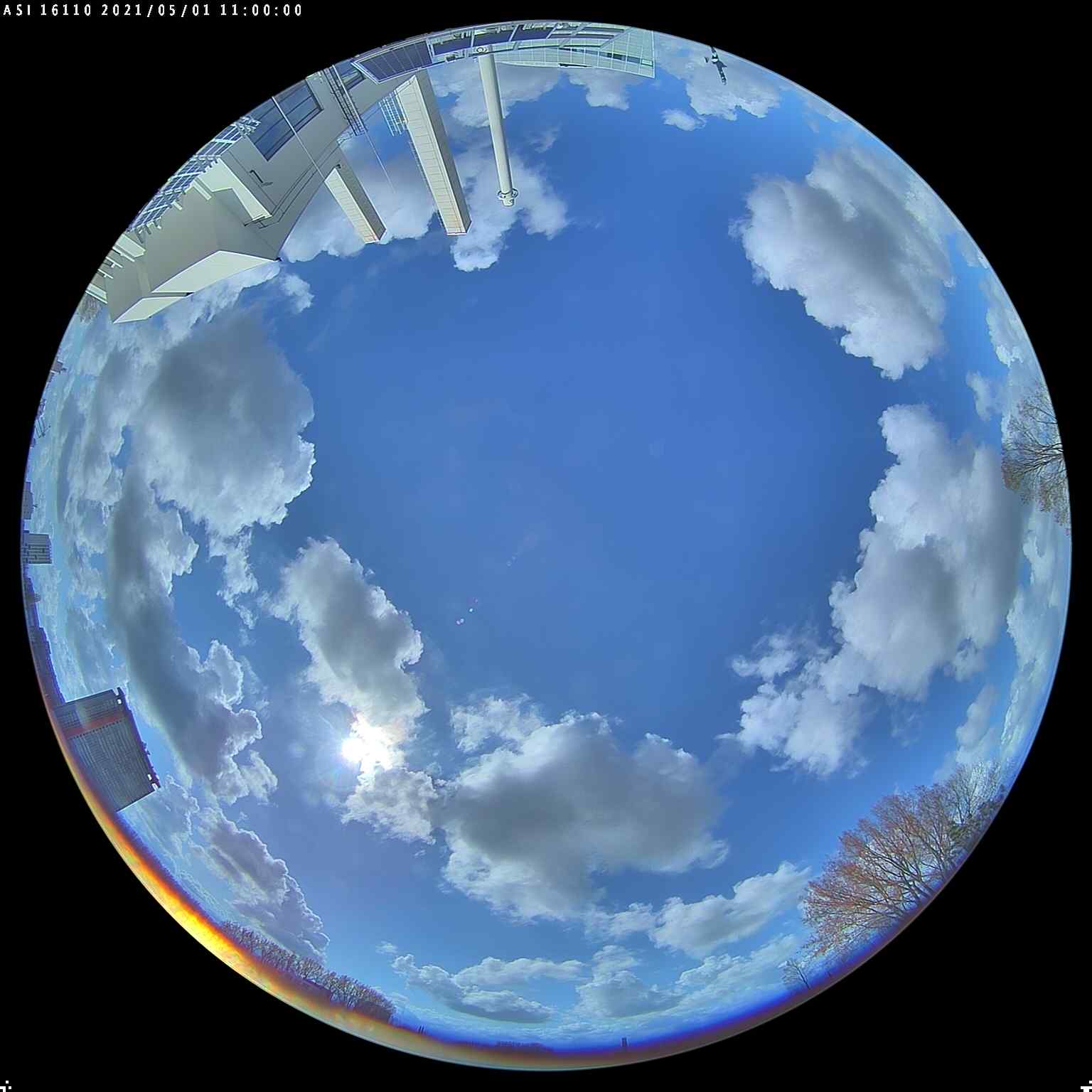}
    \caption{All sky images captured by the camera at 11:00 on 01 May 2021.}
    \label{fig:sky01may}

\end{figure}

 Sky imagery data and meteorological sensor data (with $10$ minutes resolution) are two modalities. For meteorological features, time, air temperature, cloud opacity, relative humidity, wind direction, wind speed, precipitable water, and surface pressure are considered (more details in Appendix \ref{appendix}). GHI is not included to simulate a context where local pyranometer data is not available. The sky imagery was captured at the TU Delft campus. The dataset consists of images captured at a minute frequency between 04:00 and 23:00 from 01 May $2021$ to $24$ February $2022$. Each image is in full colour, with a resolution of $1536\times1536$. An example of an image taken on $01$ May $2021$ at 11:00 is provided in Fig. \ref{fig:sky01may}.
\subsubsection{Training and Testing}
The data set contains $10,000$ data samples, where $25$\% is saved for testing, $67.5$\% is used for training (total $N=6750$ training samples), and $7.5$\% is for validation. During training, early stopping is implemented if the validation error does not decrease for 5 consecutive epochs. The AdamW optimiser with a batch size of $64$ is used to train the ANN. 
Furthermore, $10$ random train-test trials are considered for all case studies. The separation of the validation and training split between trials is varied, as is the mini-batch shuffling, while the never-before-seen test set is the same. The plots and tables in each case study show the distribution of test set predictions over the $10$ trials. Moreover, for sequential learning models (not E2E models), the mean squared error
\begin{equation}\label{eq:MSE}
MSE = \frac{1}{N} \sum_{i \in \Omega^N} (y_i - \hat{y}_i)^2
\end{equation}
and mean absolute error $MAE = \frac{1}{N}\sum_{i \in \Omega^N} |y_i - \hat{y}_i|$ were used as loss functions.

\subsubsection{Network Architecture}
Several studies were conducted to obtain a good ANN network architecture. The studies focused on the ideal ANN depth, number of images used, image resolution, etc. The \textit{CNN-base} model contains $3$ sets of convolutional and max. pooling layers, followed by $2$ FC layers; whereas the \textit{FNN-base}) for forecasting from meteorological consists of $4$ FC layers. Apart from the last layer (which uses Sigmoid activation), all other layers use the ReLU activation function. Details of these architectures are in Appendix \ref{appendix}. Moreover, these initial studies showed that using $3$ recent images, combined with a pixel-wise average, produced better PV power predictions by CNN-base than $1$ image. The same conclusion was found for the number of recent meteorological samples, resulting in $24$ meteorological inputs ($3$ for each of the $8$ variables). The initial results in Table \ref{tab:case5} show that power predictions with the meteorological data and FNN-base were far more accurate than using imagery and CNN-base. 
\begin{table}
\caption{Comparison of predictions of PV and wind power from Imagery (using CNN-base) and Meteorological data (using FNN-base).}
\label{tab:case5}
\begin{tabular}{l|cc|cc}
            & \multicolumn{2}{c|}{PV}                     & \multicolumn{2}{c}{Wind}                  \\ \hline
            & \multicolumn{1}{c|}{Imagery}  & Meteo. Data & \multicolumn{1}{c|}{Imagery} & Meteo. Data \\ \hline
Average MSE & \multicolumn{1}{c|}{0.00836}  & 0.00177     & \multicolumn{1}{c|}{0.00867} & 0.00011     \\ 
STD MSE     & \multicolumn{1}{c|}{0.000682} & 0.000134    & \multicolumn{1}{c|}{0.00131} & 0.00013     \\
\end{tabular}
\vspace{-4mm}
\end{table}
The relative benefit of meteorological data can be explained by that images do not contain as many obvious and distinct features, such as wind speed and humidity, as are directly present in the meteorological data.

\subsubsection{Investigated Models} 
 Five models were investigated.
\begin{enumerate}
	\item \textbf{UM-base1-Seq}: this naive ensemble baseline combines the predictions from imagery (using CNN-base) $f_1$ and meteorological data (using FNN-base) $f_2$ from Eq. \eqref{rawANNs} as an average
  \begin{equation}
  \hat{y}^{UM1} = 0.5 f_1(x_i^{Image}) + 0.5 f_2(x_i^{Meteo}),
 \end{equation}
 exemplifying the simplest late fusion method to combine the two modalities. The predicted power is then used to solve for system cost $C^{sys}$ in the $6$-bus system.
	\item \textbf{UM-base2-Seq}: a linear regressor is fitted to combine the two power predictions
    \begin{equation} \label{eq:ensemble}
        \hat{y}^{UM2} = c_1 f_1(x_i^{Image}) + c_2 f_2(x_i^{Meteo}),
    \end{equation}
    which can be considered as a more `intelligent' version of UM-base1-Seq as the weights $c_1$ and $c_2$ are fitted.
    \item \textbf{MM-Seq}: the MM approach shown in Fig. \ref{fig:mm4} is trained by using the $MSE$ shown in Eq. \eqref{eq:MSE} as loss, so the trained ANN has the objective to accurately predicting renewable power. This baseline represents current ML-based approaches for PV and wind prediction.
    \item \textbf{MM-E2E} (proposed model): the MM approach is used to predict power and is trained E2E based on the minimised $C^{sys}$.
    \item \textbf{Perfect Forecast}: simulates a scenario where the system cost is calculated using the true PV or wind powers $(P^{PV}_b,P^{W}_b)$. This simulation, therefore, calculates the minimum achievable system cost.
\end{enumerate}

\subsection{Meteorological Feature Extraction}
\label{case6}
Here, two MM networks are compared in terms of their effectiveness at predicting PV power. The first includes a network for extracting features from both modalities separately (FNN for meteorological data and CNN for imagery), before combining them, whereas the second only uses a network (CNN) to extract features from imagery and then fuses them with the meteorological data. The method used for fusion in both architectures for this study is concatenation. Over $10$ times training, the first model achieves an average MSE of 0.00143 (0.00058), where the number in parentheses is the standard deviation, compared to an MSE of 0.00129 (0.00031) for the second model. Despite the added complexity of the first architecture (i.e., more layers), it attains a higher error and variance in error. These results may be attributed to the fact that there are few meteorological features, to begin with ($F = 24$), thus, the added feature extraction network only overfits the training data without revealing new information from the features. This effect may be alleviated with much larger sets of training data or the use of more previous instances of meteorological features.
\subsection{Multi-Modal Joint Representation Method}
\label{case7}
This study investigates bilinear pooling (given in Eq. \eqref{eq:bipool}) and concatenation for feature fusion within the MM architecture. This fusion occurs, as shown in Fig. \ref{fig:mm4}, after feature extraction from the imagery modality. The ANN using concatenation achieves an average MSE of 0.00109 (0.00024), compared to an MSE of 0.00144 (0.00033) for bilinear pooling over these 10 trials. Bilinear pooling includes all the features represented in the fused feature space of concatenation. However, it results in quadratically larger feature space with greater sparsity and more irrelevant features. Thus, the ANN may miss important information amongst the vast feature space, potentially prompting the need for feature selection. Feature selection for regression problems reduces not only computational time but also improves predictive performance by avoiding spurious correlations between irrelevant features and the target (a form of overfitting). 

\begin{table}
	\centering
	\caption{Average and standard deviation of MSE and \%RMSE on predicted PV power over 10 trials.}
	\begin{tabular}{l|c|c|c|c}
		& \multicolumn{1}{c|}{Persistence} & \multicolumn{1}{c|}{UM-base1} & \multicolumn{1}{c|}{UM-base2} & \multicolumn{1}{c}{MM} \\ \hline
		Average MSE & 0.0008                           & 0.00294                       & 0.00151                       & 0.00109                  \\
		STD (MSE)    & 0                                & 0.000157                   & 0.000179                     & 0.000242              \\
		Average \%RMSE & 2.8\%                           & 5.4\%                       & 3.9\%                       & 3.2\%                  \\
		STD (\%RMSE) & 0                           & 0.14\%                       & 0.23\%                       & 0.34\%                  \\
	\end{tabular}
	\label{tab:case8}

\vspace{-4mm}
\end{table}
\vspace{-1mm}
\subsection{Comparison of Methods for Power Forecasting}
\label{case8}
This case study quantifies the utility of MM learning for power forecasting. UM-base1, UM-base2, the proposed MM model, and a persistence forecast are compared in terms of prediction error. The OPF problems, system cost and E2E training are therefore not considered here as we seek to compare the prediction performance of the models. The results are presented in Table \ref{tab:case8}. The prediction performance of UM-base1, UM-base2 and MM, in terms of \%RMSE, is in line with state-of-the-art ANN-based methods for predicting solar power \cite{Raza2016}. The persistence forecast outperforms the other $3$ methods consistently. However, it must be noted that this is because it is predicting PV power (analogous to GHI given Eq. \eqref{eq:pvoutput}) using the GHI from 10 minutes prior, while the ANN methods do not include recent GHI as a feature. UM-base1 severely lacks performance compared to UM-base2, likely due to the significantly worse-performing imagery modality being equally weighed with the sensor data. Additionally, MM learning improves predictions compared to UM-base2. Two reasons may explain this phenomenon. Firstly, UM-base2 has only two trainable parameters (as given in Eq. \eqref{eq:ensemble}), while MM has thousands; this additional complexity may aid prediction. Secondly, UM-base2 employs the latest possible fusion, only combining the predicted GHI from either modality, while MM employs intermediate fusion, concatenating features from meteorological data and imagery. During training, this architecture extracted feature layers (CNN-fe) to learn how the visual features enhance the sensor data for forecasting.
\vspace{-1mm}
\subsection{E2E learning with PV and Wind Prediction}
\label{case9}
\begin{figure}
	\centering
	\includegraphics[width=0.37\textwidth]{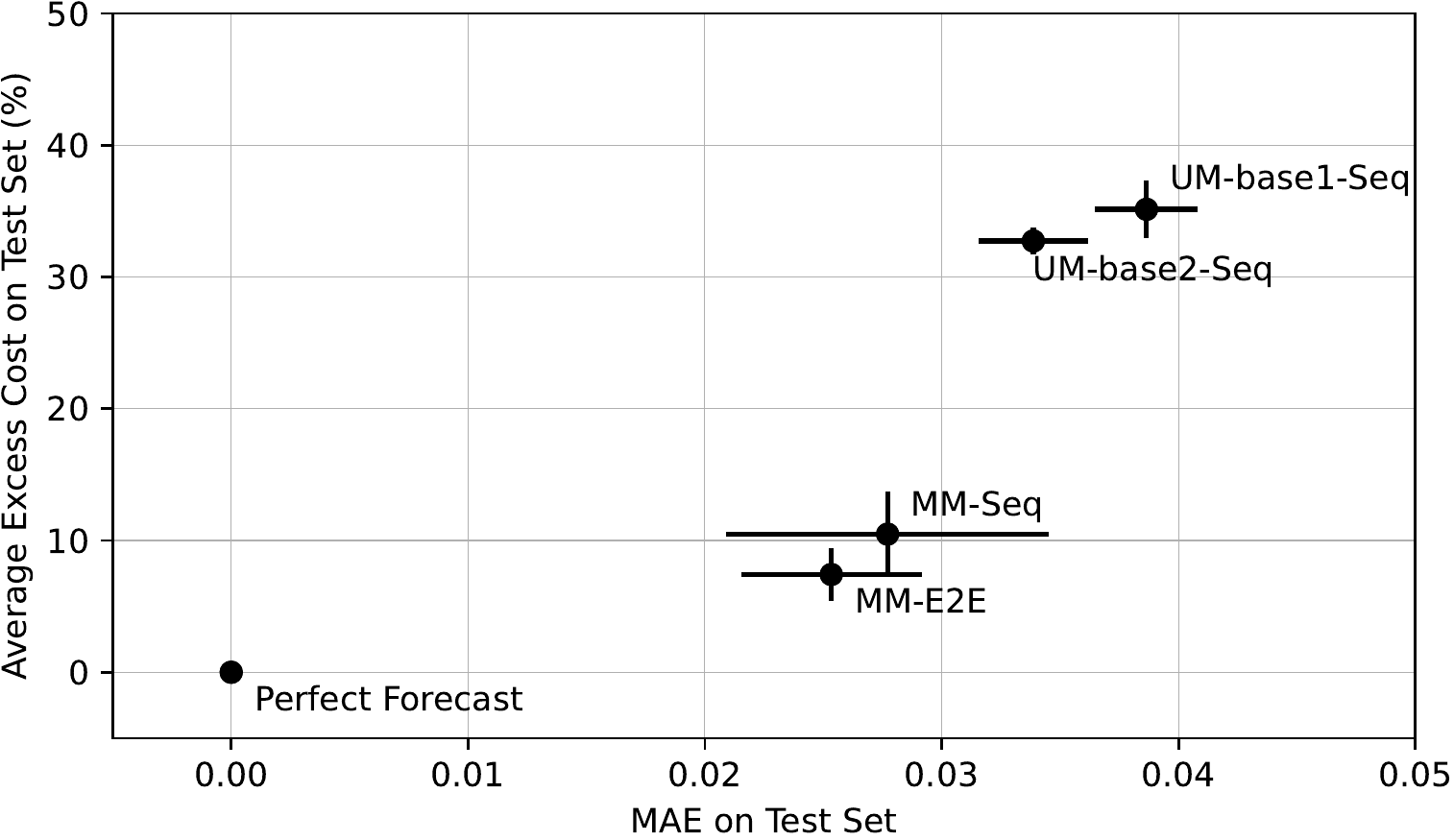}
	\caption{Excess system cost compared to power forecast error (MAE) in a system with PV generation.}
	\label{fig:case9}
\vspace{-4mm}
\end{figure}
This study investigates the system costs minimisation by E2E learning against the baselines for predicting PV power which are established already. The MM learning model in Fig. \ref{fig:case9} is shown to not only improve power output prediction but also benefit power system operation in terms of system cost as both the MM models outperform the UM ones. A conventional approach, represented by UMbase2-Seq, results in a 32\% cost increase compared to a perfect knowledge scenario. $C^{sys}$ is only increased by 10\% when using MM learning and training sequentially. Much of this cost reduction is due to the intelligent combination of MM features using an ANN. E2E learning also contributes a clear reduction in cost (comparing the 10\% excess cost from MM-Seq to 7\% from MM-E2E). In addition, E2E training enables a more consistently reduced cost, shown by the lower variance in cost. This improved performance of MM-E2E can be explained by the fact that E2E training provides the ANN with information about the generation cost functions ($c_b^{up}$, $c_b^{dn}$) of the various generators. Thus, for instances where perfect prediction is not obtained, E2E training may enable the ANN to appropriately over- or underestimate power such to reduce cost. 
\par
\begin{figure}
	\centering
	\includegraphics[width=0.37\textwidth]{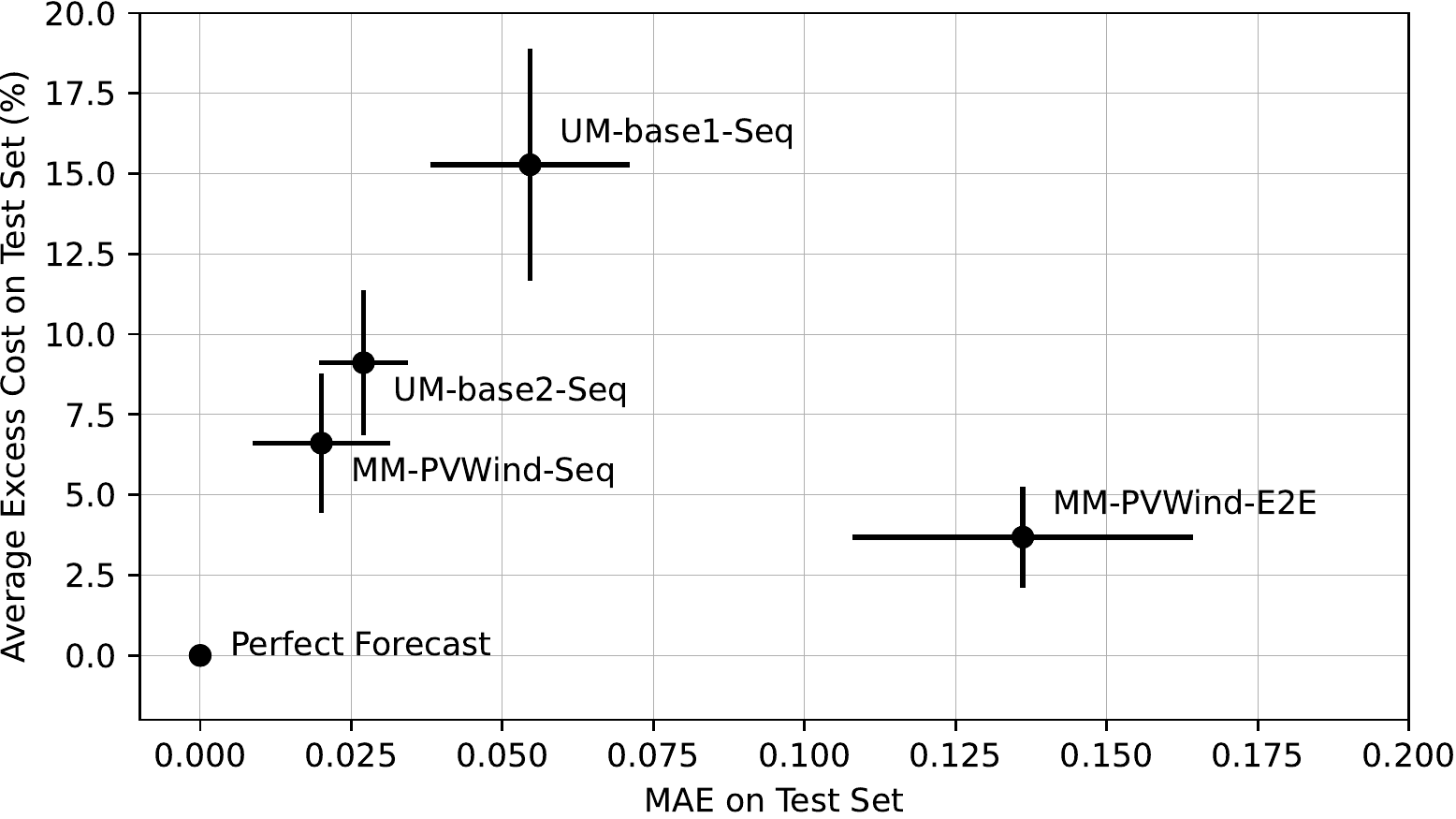}
	\caption{Excess system cost compared to power forecast error (MAE) in a system with PV and wind generation.}
	\label{fig:case10}
 \vspace{-7mm}
\end{figure}

 This same study can be conducted in a power system with PV and wind generation. The proposed model is adjusted to `MM-PVWind-E2E' to predict both power sources by including an FNN-pred network which uses only meteorological data for wind prediction. The results are in Fig. \ref{fig:case10}, where the error is measured as the average of $MAE(\hat{P}_g^{PV},P_g^{PV})$ and $MAE(\hat{P}_g^{W},P_g^{W})$.
While E2E training results in lower system costs than sequential learning, the power prediction error is significantly greater in the former. Note that the standard deviation on system cost is lowest from MM-PVWind-E2E, while the standard deviation on the power prediction is the largest for MM-PVWind-E2E.
\vspace{-1mm}
\subsection{System Cost Functions}
\label{case11}
\begin{figure}
    \centering
    \begin{subfigure}[b]{0.24\textwidth}
         \centering
         \includegraphics[width=\textwidth]{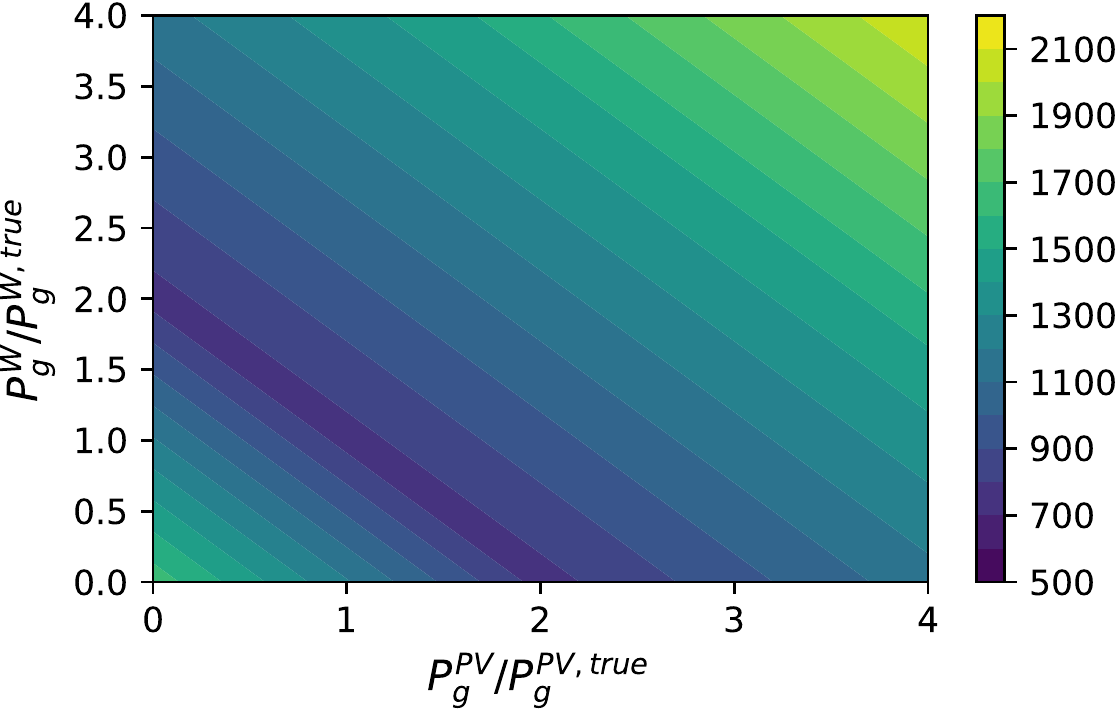}
         \caption{}
         \label{fig:system21}
     \end{subfigure}
     \hfill
     \begin{subfigure}[b]{0.24\textwidth}
         \centering
         \includegraphics[width=\textwidth]{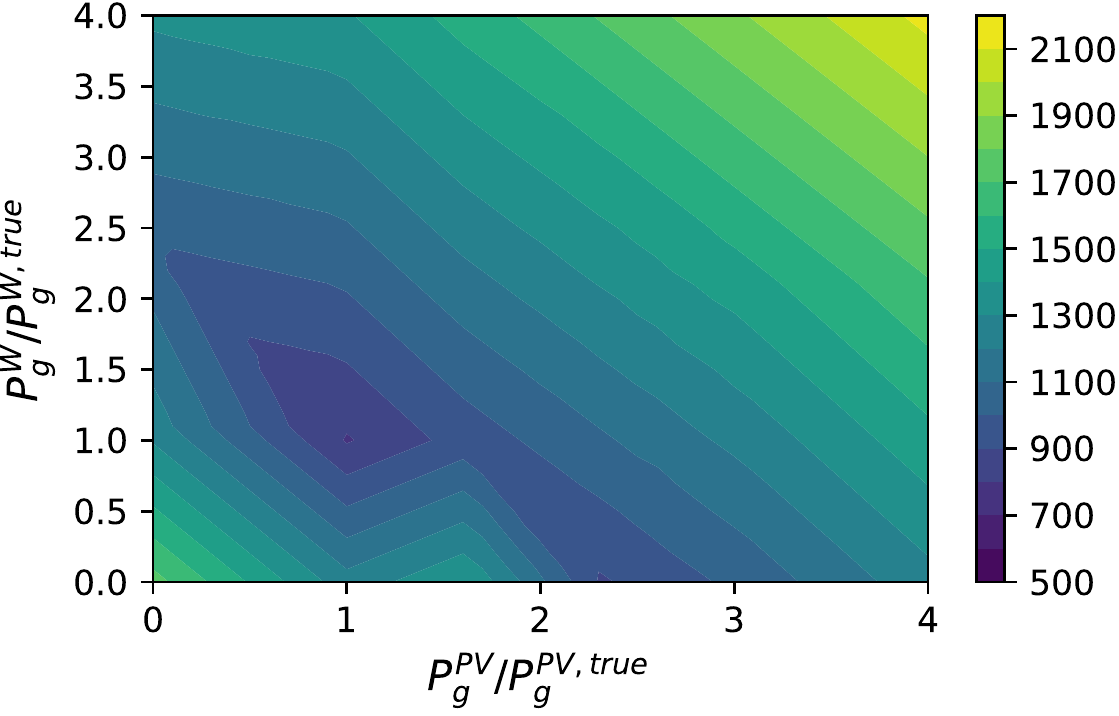}
         \caption{}
         \label{fig:system22}
     \end{subfigure}
     \caption{Cost functions of a system with PV and wind generation, (a) under regular, (b) and reduced (by 50\%) branch limits.}
     \label{fig:system2}
\vspace{-4mm}
\end{figure}

The results found in Figs. \ref{fig:case9} and \ref{fig:case10} are worth investigating as they both show MM-E2E achieving the lowest costs but with disparate prediction error. We investigate the cost function in PV and wind systems with a load of 145MW, and PV and wind generation of $25$MW each. Fig. \ref{fig:system2} shows the system cost $C^{sys}$ under regular and reduced (by $50$\%) branch limits. Fig. \ref{fig:system21} shows no global minimum, but isolines along which identical costs can be achieved by overestimating PV and underestimating wind, or vice-versa. Thus, system balance may be maintained with no (or very low) redispatch costs while incurring prediction errors. 
This behaviour explains Fig. \ref{fig:case10}, where relatively large prediction errors are not detrimental to cost minimisation. Fig. \ref{fig:system22} shows possible local minima as branch limits are reduced and more constraints are binding. E2E training in this context would enable the learning of a more complex cost function, but may also cause inaccurate power predictions. Thus, the optimisation objective should be carefully selected depending on the problem context for E2E training.
\vspace{-2mm}
\subsection{Discussion}
\label{sec:limitations}
The studies presented promising results regarding the application of MM and E2E learning for economically optimised renewable power predictions. However, some limitations of the methodology can be noted, which most importantly is the computational time to train the models as it involves solving optimisations. The computational time's complexity to the training data and power network sizes must be thoroughly analysed to quantify the approach's limitations. Furthermore, in future works, using Pearson correlation for selecting features may improve power prediction. 
Including features from early layers into the MM power prediction may also improve performance and mitigate the vanishing gradients \cite{Baltrusaitis2019}. Testing the proposed MM-E2E architecture with various cost functions, e.g., asymmetric and quadratic ones would simulate more realistic conditions. Designing a tailored loss function, such as a multi-task objective, may prevent converging to local minima (sometimes seen with E2E learning). Further, considering the prediction relies mostly on time-sequenced data (in both modalities), ML techniques such as LSTM-RNNs could also improve predictions compared to only CNNs and FNNs. Finally, a more thorough analysis with a wider set of OPF-type problems is recommended to fully assess the proposed approach's benefits.
\section{Conclusion}
\label{sec:conc}
This paper investigated two deep learning techniques, multi-modal (MM) learning and end-to-end (E2E) learning, for photovoltaic and wind power nowcasting in energy management systems. The features from two modalities (all-sky imagery and time series meteorological data) are combined through MM learning. The model is trained E2E on the total system cost from prediction errors modelled as optimization problems. The proposed model architecture, \textit{MM-E2E}, is tested in different settings against uni-modal sequentially trained convolutional and feedforward networks to understand the advantages and limitations of the methodology. Our studies found that the proposed \textit{MM-E2E} model reduced the economic cost by impressively around 30\%. However, we are also cautious and discussed the limitation regarding the computational scaling of the E2E  approach, which we recommend as future works.

\bibliographystyle{IEEEtran}
\bibliography{conference}

\begin{figure}[h]
	\centering
	\includegraphics[width=0.45\textwidth]{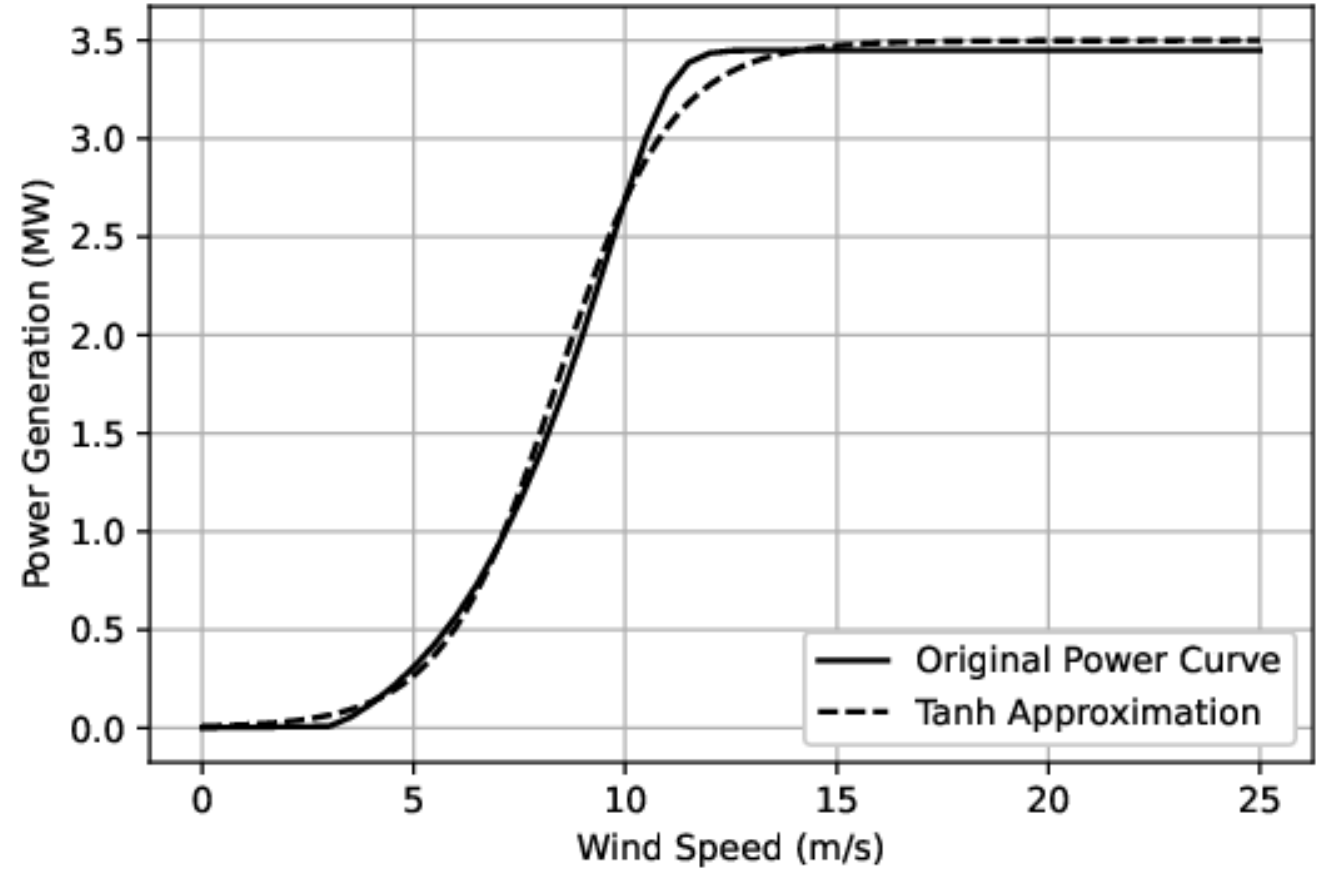}
	\caption{Power curve of Vestas V112-3.45 turbine and a Tanh approximation.}
	\label{fig:turbine}
\end{figure}

\begin{appendix}
\label{appendix}
Details of the parameters used in the DCOPF-Schedule and DCOPF-Redispatch optimisations are given in Table \ref{tab:dcopf_params}. The wind turbine power curve and its approximation are depicted in Fig. \ref{fig:turbine}. 
This work considers meteorological sensor data and sky imagery data modalities. The meteorological features are presented in Table \ref{tab:meteo_feat}. Each one is measured at a temporal resolution of $10$ minutes. The baseline models resulting from initial case studies (CNN-base and FNN-base) are described in Tables \ref{tab:cnn-base} and \ref{tab:fnn-base}. The architectures of CNN-fe and FNN-pred are given in Tables \ref{tab:cnn-fe} and \ref{tab:fnn-pred}.

\begin{table}[h]
	\centering
	\caption{Values of essential parameters used in the optimisations.}
	\label{tab:dcopf_params}
	\begin{tabular}{c|c|c}
		Parameter             & Value & Unit    \\ \hline
		$\underline{P^g_1}, \underline{P^g_2}, \underline{P^g_3}$            & 0     & MW      \\[0.15cm] 
		$\overline{P^g_1}$            & 200   & MW      \\[0.15cm]
		
		$\overline{P^g_2}$            & 150   & MW      \\[0.15cm]
		
		$\overline{P^g_2}$            & 180   & MW      \\[0.15cm]
		$c_1^g$               & 12    & Euro/MW \\[0.1cm]
		$c_2^g$               & 10    & Euro/MW \\[0.1cm]
		$c_3^g$               & 8     & Euro/MW \\[0.1cm]
		$\overline{f_l}$         & 100   & MW      \\
		$L$                   & 11    &         \\
		$B$                   & 6     &         \\
		$\gamma_b$      & 100    & Euro/MW \\[0.1cm]
		$\gamma^{PV}_b, \gamma^{W}_b$      & 0.1    & Euro/MW \\
	\end{tabular}
\end{table}

\begin{table}
	\centering
	\caption{FNN-base architecture.}
	\label{tab:fnn-base}
	\begin{tabular}{|llllc|}
		\hline
		\multicolumn{1}{|c|}{Layer} & \multicolumn{1}{c|}{Type}            & \multicolumn{1}{c|}{Input Size}                                  & \multicolumn{1}{c|}{\# Neurons}        &  \multicolumn{1}{c|}{Activation}                                 \\ \hline
		\multicolumn{1}{|c|}{F1}    & \multicolumn{1}{c|}{FC} & \multicolumn{1}{c|}{$n_{inputs}\times$1}                                & \multicolumn{1}{c|}{64}                       &  \multicolumn{1}{c|}{ReLU}                                  \\
		\multicolumn{1}{|c|}{F2}    & \multicolumn{1}{c|}{FC} & \multicolumn{1}{c|}{64$\times$1}                                & \multicolumn{1}{c|}{128}                       &\multicolumn{1}{c|}{ReLU}                                  \\
		\multicolumn{1}{|c|}{F3}    & \multicolumn{1}{c|}{FC} & \multicolumn{1}{c|}{128$\times$1}                                & \multicolumn{1}{c|}{64}                       &  \multicolumn{1}{c|}{ReLU}                                  \\
		\multicolumn{1}{|c|}{F4}    & \multicolumn{1}{c|}{FC} & \multicolumn{1}{c|}{64$\times$1}                                 & \multicolumn{1}{c|}{1}                        &  \multicolumn{1}{c|}{Sigmoid}
		\\ \hline                              
	\end{tabular}
\end{table}

\begin{table}
	\centering
	\caption{CNN-base architecture.}
	\label{tab:cnn-base}
	\begin{tabular}{|llllllc|}
		\hline
		\multicolumn{1}{|c|}{Layer} & \multicolumn{1}{c|}{Type}            & \multicolumn{1}{c|}{Input Size} & \multicolumn{1}{c|}{Kernel} & \multicolumn{1}{c|}{Stride } & \multicolumn{1}{c|}{Padding } & \multicolumn{1}{c|}{Activation} \\ \hline
		\multicolumn{1}{|c|}{C1}    & \multicolumn{1}{c|}{Conv.}   & \multicolumn{1}{c|}{64$\times$64$\times$1}                       & \multicolumn{1}{c|}{3$\times$3}               & \multicolumn{1}{c|}{2$\times$2}          & \multicolumn{1}{c|}{1$\times$1}           & \multicolumn{1}{c|}{ReLU}                       \\
		\multicolumn{1}{|c|}{P2}    & \multicolumn{1}{c|}{Max. Pool.}    & \multicolumn{1}{c|}{32$\times$32$\times$8}                       & \multicolumn{1}{c|}{2$\times$2}               & \multicolumn{1}{c|}{2$\times$2}          & \multicolumn{1}{c|}{0$\times$0}           & \multicolumn{1}{c|}{-}                       \\
		\multicolumn{1}{|c|}{C3}    & \multicolumn{1}{c|}{Conv.}   & \multicolumn{1}{c|}{16$\times$16$\times$8}                       & \multicolumn{1}{c|}{3$\times$3}               & \multicolumn{1}{c|}{1$\times$1}          & \multicolumn{1}{c|}{1$\times$1}           & \multicolumn{1}{c|}{ReLU}                       \\
		\multicolumn{1}{|c|}{P4}    & \multicolumn{1}{c|}{Max. Pool.}    & \multicolumn{1}{c|}{16$\times$16$\times$32}                      & \multicolumn{1}{c|}{2$\times$2}               & \multicolumn{1}{c|}{2$\times$2}          & \multicolumn{1}{c|}{0$\times$0}           & \multicolumn{1}{c|}{-}                         \\
		\multicolumn{1}{|c|}{C5}    & \multicolumn{1}{c|}{Conv.}   & \multicolumn{1}{c|}{8$\times$8$\times$32}                        & \multicolumn{1}{c|}{3$\times$3}               & \multicolumn{1}{c|}{1$\times$1}          & \multicolumn{1}{c|}{1$\times$1}           & \multicolumn{1}{c|}{ReLU}                         \\
		\multicolumn{1}{|c|}{P6}    & \multicolumn{1}{c|}{Max. Pool.}    & \multicolumn{1}{c|}{8$\times$8$\times$16}                        & \multicolumn{1}{c|}{2$\times$2}               & \multicolumn{1}{c|}{2$\times$2}          & \multicolumn{1}{c|}{0$\times$0}           & \multicolumn{1}{c|}{-}                          \\ \hline
		\multicolumn{7}{|c|}{Flatten}              \\ \hline
		\multicolumn{1}{|c|}{Layer} & \multicolumn{1}{c|}{Type}            & \multicolumn{1}{c|}{Input Size}                                  & \multicolumn{1}{c|}{\# Neurons}        & \multicolumn{1}{c|}{-}                   & \multicolumn{1}{c|}{-}                    & \multicolumn{1}{c|}{Activation}                                  \\ \hline
		\multicolumn{1}{|c|}{F7}    & \multicolumn{1}{c|}{FC} & \multicolumn{1}{c|}{256$\times$1}                                & \multicolumn{1}{c|}{64}                       & \multicolumn{1}{c|}{-}                   & \multicolumn{1}{c|}{-}                    & \multicolumn{1}{c|}{ReLU}                               \\
		\multicolumn{1}{|c|}{F8}    & \multicolumn{1}{c|}{FC} & \multicolumn{1}{c|}{64$\times$1}                                 & \multicolumn{1}{c|}{1}                        & \multicolumn{1}{c|}{-}                   & \multicolumn{1}{c|}{-}                    & \multicolumn{1}{c|}{Sigmoid}
		\\ \hline                              
	\end{tabular}
\end{table}

\begin{table}
	\centering
	\caption{CNN-fe architecture.}
	\label{tab:cnn-fe}
	\begin{tabular}{|llllllc|}
		\hline
		\multicolumn{1}{|c|}{Layer} & \multicolumn{1}{c|}{Type}            & \multicolumn{1}{c|}{Input Size} & \multicolumn{1}{c|}{Kernel} & \multicolumn{1}{c|}{Stride } & \multicolumn{1}{c|}{Padding } & \multicolumn{1}{c|}{Activation} \\ \hline
		\multicolumn{1}{|c|}{C1}    & \multicolumn{1}{c|}{Conv.}   & \multicolumn{1}{c|}{64$\times$64$\times$1}                       & \multicolumn{1}{c|}{3$\times$3}               & \multicolumn{1}{c|}{2$\times$2}          & \multicolumn{1}{c|}{1$\times$1}           & \multicolumn{1}{c|}{ReLU}                     \\
		\multicolumn{1}{|c|}{P2}    & \multicolumn{1}{c|}{Max. Pool.}    & \multicolumn{1}{c|}{32$\times$32$\times$8}                       & \multicolumn{1}{c|}{2$\times$2}               & \multicolumn{1}{c|}{2$\times$2}          & \multicolumn{1}{c|}{0$\times$0}           & \multicolumn{1}{c|}{-}                \\
		\multicolumn{1}{|c|}{C3}    & \multicolumn{1}{c|}{Conv.}   & \multicolumn{1}{c|}{16$\times$16$\times$8}                       & \multicolumn{1}{c|}{3$\times$3}               & \multicolumn{1}{c|}{1$\times$1}          & \multicolumn{1}{c|}{1$\times$1}           & \multicolumn{1}{c|}{ReLU}                \\
		\multicolumn{1}{|c|}{P4}    & \multicolumn{1}{c|}{Max. Pool.}    & \multicolumn{1}{c|}{16$\times$16$\times$32}                      & \multicolumn{1}{c|}{2$\times$2}               & \multicolumn{1}{c|}{2$\times$2}          & \multicolumn{1}{c|}{0$\times$0}           & \multicolumn{1}{c|}{-}                       \\
		\multicolumn{1}{|c|}{C5}    & \multicolumn{1}{c|}{Conv.}   & \multicolumn{1}{c|}{8$\times$8$\times$32}                        & \multicolumn{1}{c|}{3$\times$3}               & \multicolumn{1}{c|}{1$\times$1}          & \multicolumn{1}{c|}{1$\times$1}           & \multicolumn{1}{c|}{ReLU}                        \\
		\multicolumn{1}{|c|}{P6}    & \multicolumn{1}{c|}{Max. Pool.}    & \multicolumn{1}{c|}{8$\times$8$\times$16}                        & \multicolumn{1}{c|}{2$\times$2}               & \multicolumn{1}{c|}{2$\times$2}          & \multicolumn{1}{c|}{0$\times$0}           & \multicolumn{1}{c|}{-}                         \\ \hline
		\multicolumn{7}{|c|}{Flatten}              \\ \hline
		\multicolumn{1}{|c|}{Layer} & \multicolumn{1}{c|}{Type}            & \multicolumn{1}{c|}{Input Size}                                  & \multicolumn{1}{c|}{\# Neurons}        & \multicolumn{1}{c|}{-}                   & \multicolumn{1}{c|}{-}                    & \multicolumn{1}{c|}{Activation}                                  \\ \hline
		\multicolumn{1}{|c|}{F7}    & \multicolumn{1}{c|}{FC} & \multicolumn{1}{c|}{256$\times$1}                                & \multicolumn{1}{c|}{64}                       & \multicolumn{1}{c|}{-}                   & \multicolumn{1}{c|}{-}                    & \multicolumn{1}{c|}{ReLU}                                  \\
		\multicolumn{1}{|c|}{F8}    & \multicolumn{1}{c|}{FC} & \multicolumn{1}{c|}{64$\times$1}                                 & \multicolumn{1}{c|}{64}                        & \multicolumn{1}{c|}{-}                   & \multicolumn{1}{c|}{-}                    & \multicolumn{1}{c|}{Sigmoid}
		\\ \hline                              
	\end{tabular}
\end{table}
\begin{table}
	\centering
	\caption{FNN-pred architecture.}
	\label{tab:fnn-pred}
	\begin{tabular}{|llllc|}
		\hline
		\multicolumn{1}{|c|}{Layer} & \multicolumn{1}{c|}{Type}            & \multicolumn{1}{c|}{Input Size}                                  & \multicolumn{1}{c|}{\# Neurons}        &  \multicolumn{1}{c|}{Activation}                            \\ \hline
		\multicolumn{1}{|c|}{F1}    & \multicolumn{1}{c|}{FC} & \multicolumn{1}{c|}{88$\times$1}                                & \multicolumn{1}{c|}{64}                       &  \multicolumn{1}{c|}{ReLU}                                  \\
		\multicolumn{1}{|c|}{F2}    & \multicolumn{1}{c|}{FC} & \multicolumn{1}{c|}{64$\times$1}                                & \multicolumn{1}{c|}{128}                       &\multicolumn{1}{c|}{ReLU}                                       \\
		\multicolumn{1}{|c|}{F3}    & \multicolumn{1}{c|}{FC} & \multicolumn{1}{c|}{128$\times$1}                                & \multicolumn{1}{c|}{64}                       &  \multicolumn{1}{c|}{ReLU}                                   \\
		\multicolumn{1}{|c|}{F4}    & \multicolumn{1}{c|}{FC} & \multicolumn{1}{c|}{64$\times$1}                                 & \multicolumn{1}{c|}{1}                        &  \multicolumn{1}{c|}{Sigmoid}   
		\\ \hline                              
	\end{tabular}
\end{table}

\begin{table}
	\centering
	\caption{Meteorological features used for predicting $P^{PV}$ or $P^{W}$.}
	\label{tab:meteo_feat}
	\begin{tabular}{p{0.12\textwidth}|p{0.21\textwidth}|p{0.07\textwidth}}
		Meteo. Variable & \hfil Description           & \hfil Unit       \\ \hline
		Time                    & The moment of the day when the instance was measured  & seconds    \\
		Air temperature         & Measured at 2 meters above surface level              & $\degree$C \\
		Cloud opacity           & The attenuation of incoming sunlight due to clouds. Varies from 0\% (no cloud) to 100\% (full attenuation of incoming sunlight)                               & \%         \\
		Relative humidity       & Measured at 2 meters above ground level. Relative humidity is the amount of water vapour as a percentage of the amount needed for saturation at the same temperature. A value of 50\% means the air is 50\% saturated & \%         \\
		Wind direction          & Measured at 10m altitude                            & $\degree$  \\
		Wind speed              & Measured at 10m altitude                            & m/s        \\
		Precipitable water      & A measure of the precipitable water of the entire atmospheric column                      & kg/m$^2$   \\
		Surface pressure        & Air pressure measured at sea level                   & hPa       
	\end{tabular}
\end{table}
\end{appendix}
\end{document}